\documentclass[aps,pre,showpacs,amsmath,amssymb,amsfonts,superscriptaddress,twocolumn,lengthcheck]{revtex4-1}

\usepackage{graphicx}
\usepackage{subfigure}
\usepackage{verbatim}
\usepackage{dcolumn}
\usepackage{bm}
\usepackage{epsf}
\usepackage{color}
\usepackage[colorlinks=true,citecolor=blue,linkcolor=blue,urlcolor=blue]{hyperref}
\usepackage{hhline}

\newcommand{\tr}[1]{\mathrm{tr}\left\{#1\right\}}

\newcommand{\la}{\left\langle}
\newcommand{\ra}{\right\rangle}
\newcommand{\pd}{\partial}

\newcommand{\td}{\mathrm{d}}

\newcommand{\e}[1]{\exp{\left(#1\right)}}

\newcommand{\com}[2]{\left[#1,\,#2\right]}

\newcommand{\co}[1]{\cos{\left(#1\right)}}
\newcommand{\si}[1]{\sin{\left(#1\right)}}

\newcommand{\bla}{bla\\bla\\bla\\bla\\bla}

\newcommand{\PRA}{Phys. Rev. A }

\newcommand{\PRE}{Phys. Rev. E }
\newcommand{\PRL}{Phys. Rev. Lett. }

\newcommand{\RMP}{Rev. Mod. Phys. }

\newcommand{\mc}[1]{\mathcal{#1}}

\newcommand{\mrm}[1]{\mathrm{#1}}

\begin{document}

\title{Shortcuts to adiabaticity from linear response theory}

\author{Thiago V. Acconcia} 
\email[]{thiagova@ifi.unicamp.br}
\affiliation{Instituto de F\'\i sica  `Gleb Wataghin', Universidade Estadual de Campinas - Unicamp,  Rua S\'ergio Buarque de Holanda 777,  13083-859 Campinas, SP, Brazil}
  
\author{Marcus V. S. Bonan\c{c}a}
\affiliation{Instituto de F\'\i sica  `Gleb Wataghin', Universidade Estadual de Campinas - Unicamp,  Rua S\'ergio Buarque de Holanda 777,  13083-859 Campinas, SP, Brazil}
  
\author{Sebastian Deffner}
\affiliation{Theoretical Division and Center for Nonlinear Studies, Los Alamos National Laboratory, Los Alamos, New Mexico 87545, USA}


\date{\today}

\begin{abstract} 
A shortcut to adiabaticity is a finite-time process that produces the same final state as would result from infinitely slow driving. We show that such shortcuts can be found for weak perturbations from linear response theory. With the help of phenomenological response functions, a simple expression for the excess work is found--quantifying the nonequilibrium excitations. For two specific examples, i.e., the quantum parametric oscillator and the spin 1/2 in a time-dependent magnetic field, we show that finite-time zeros of the excess work indicate the existence of shortcuts. Finally, we propose a degenerate family of protocols, which facilitates shortcuts to adiabaticity for specific and very short driving times.
\end{abstract}

\maketitle

\section{Introduction}

Thermodynamics is a phenomenological theory to describe the transformation of heat into work. However, only  quasistatic, i.e., infinitely slow processes  are fully describable by means of conventional thermodynamics \cite{callen}.  For all realistic, finite-time--\emph{nonequilibrium}--processes, the second law of thermodynamics constitutes merely an \emph{inequality}, expressing that some portion of the energy or entropy is irreversibly lost into nonequilibrium excitations. For isothermal processes, this ``loss'' is quantified by the excess work $\la W_\mrm{ex}\ra$, which is the difference between the total nonequilibrium work $\la W\ra$ and the work performed during a quasistatic--\emph{equilibrium}--process $\la W_\mrm{qs}\ra$, $\la W_\mrm{ex}\ra = \la W \ra - \la W_\mrm{qs} \ra$. For macroscopic, open systems, $\la W_\mrm{qs} \ra$ is simply given by the free energy difference $\Delta F$. However, the identification of the equilibrium work, $\la W_\mrm{qs}\ra$, with the free energy difference, $\Delta F$, is only true for open systems. For isolated systems, the minimal work is not given by the free energy difference and $\la W_\mrm{qs}\ra$ has to be analyzed carefully \cite{Allahverdyan2005a}. In addition, for quantum systems, the situation is particularly involved as quantum work is not an observable in the usual sense, as there is no Hermitian operator, whose eigenvalues are given by the classical work values \cite{tasaki_2000,kurchan_2000,talkner_2007,campisi_2011,Hanggi2015}.

Nevertheless, finding ``optimal'' quantum processes, for which only the minimal amount of $\la W_\mrm{ex}\ra$ is lost into nonequilibrium excitations, is of fundamental importance.  Consequently,  a lot of theoretical and experimental research has been dedicated to the  design of so-called shortcuts to adiabaticity, i.e., finite-time processes with suppressed nonequilibrium excitations \cite{torrontegui_2013}. To this end, a variety of techniques has been proposed: the use of dynamical invariants \cite{chen_2010}, the inversion of scaling laws \cite{campo_boshier_2012}, the fast-forward technique \cite{masuda_2011,torrontegui_2012}, and transitionless quantum driving \cite{demirplak_rice_2003,demirplak_rice_2005,berry_2009,Deffner2014}. All methods have in common that practical implementations are rather involved as the full dynamics has to be solved to determine the shortcut. Therefore, more recent research efforts have been focusing on identifying optimal protocols from optimal control theory \cite{Stefanatos2013,Campbell2014}, properties of the quantum work statistics  \cite{Xiao2014}, or ``environment'' assisted methods \cite{Masuda2014}. 
\begin{figure}
\centering
\includegraphics[width=.48\textwidth]{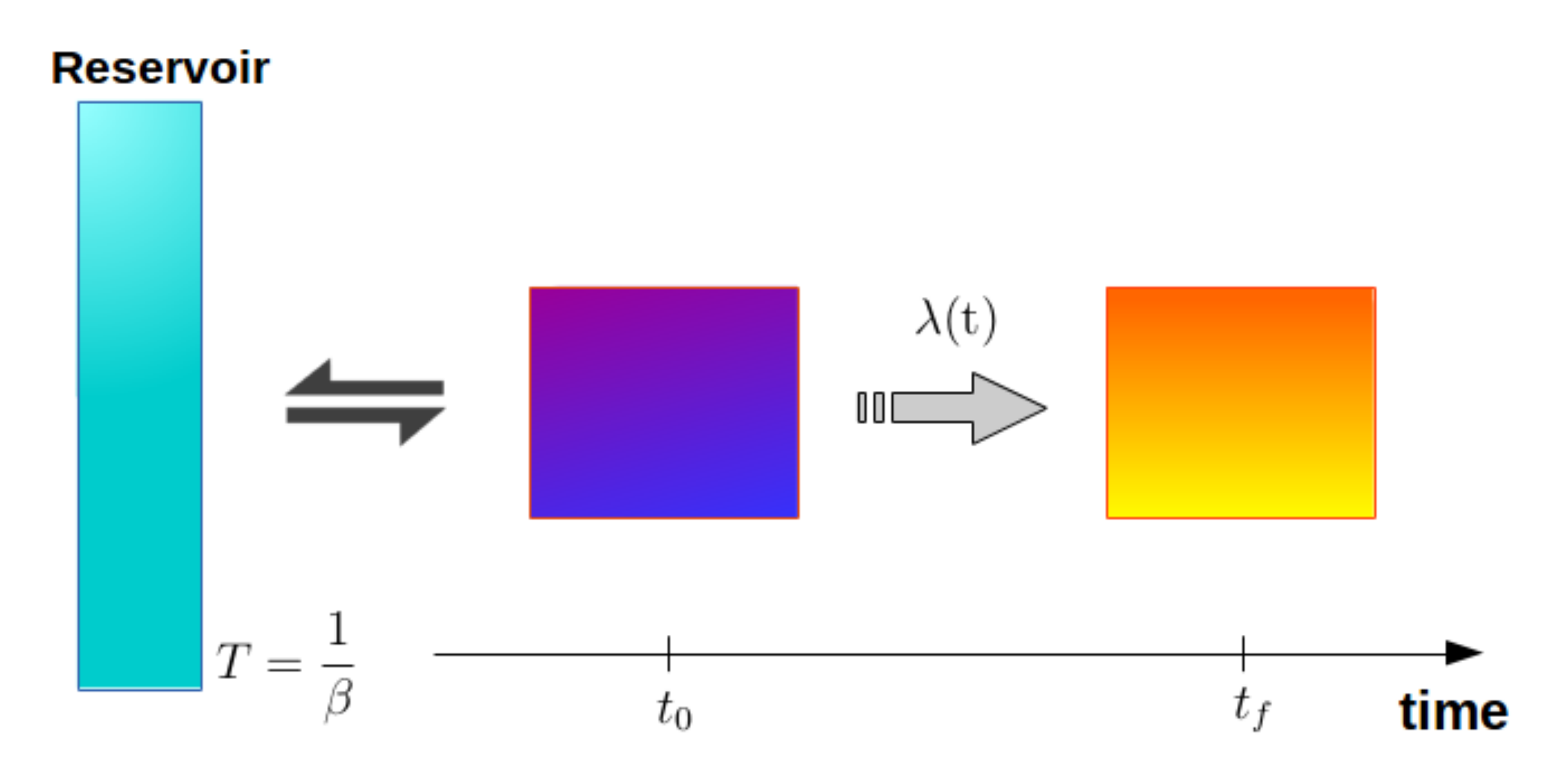} 
\caption{(Color online) Sketch of the thermodynamic processes under study. At $t=t_0$, the system is prepared in equilibrium with inverse temperature $\beta$ before the system is decoupled from the environment and controlled externally from $t=t_0+0^+$ until a final time $t_{f}$.}
\label{sketch_perturbation}
\end{figure}

The present analysis is dedicated to finding shortcuts to adiabaticity from a phenomenological approach--linear response theory. For classical systems, it has been recently shown that there exist finite-time processes with zero excess work \cite{Acconcia2014}. In this paradigm, $\la W_\mrm{ex}\ra$ is fully determined by the phenomenological response of the system to an external perturbation \cite{Sivak,Bonanca2014a}. Thus, we neither have to solve the dynamics \cite{demirplak_rice_2003,demirplak_rice_2005,berry_2009,Deffner2014} nor do we have to determine the quantum work statistics  \cite{Xiao2014}  to minimize $\la W_\mrm{ex}\ra$. In the following, we will extend our previous findings \cite{Bonanca2014a,Acconcia2014} to the quantum domain. To this end, we will consider a thermally isolated quantum system under weak perturbation and derive a linear response expression for $\la W_\mrm{ex}\ra$. After establishing the general theory, we will turn to analytically solvable and pedagogically elucidating examples, namely the parametric harmonic oscillator and the spin 1/2 in a time-dependent magnetic field. This will allow us to study the range of validity of the linear response approach by comparing our findings with the exact results from the full quantum work statistics \cite{Deffner2008,Deffner2010}. We will show that the protocols with zero excess work from linear response theory, indeed, they facilitate transitionless quantum driving for weak perturbations. Finally, we will propose  a family of degenerate protocols, which facilitates shortcuts to adiabaticity for arbitrarily fast driving.

\section{\label{sec:LRT}Quantum work from linear response theory}

We begin by generalizing the previous classical treatment of the  excess work $\la W_\mrm{ex}\ra$ \cite{Acconcia2014} to the quantum domain. Imagine a quantum system with time-dependent Hamiltonian $H_t$, which is prepared initially in a thermal equilibrium state, $\rho_0 = \exp(-\beta H_{0})/ Z_0$, where $Z_0$ is the partition function, $Z_0=\tr{\e{-\beta H_0}}$. At $t=t_0+0^+$, the system is decoupled from the environment and the Hamiltonian is varied according to some protocol $\lambda_t$ with $H_t\equiv H(\lambda_t)$. Such a processes is sketched in Fig.~\ref{sketch_perturbation}.

The external control parameter $\lambda_t$ is written as
\begin{equation}
\label{eq01}
\lambda_t \equiv \lambda_{0} + \delta\lambda \ g(t),
\end{equation}
where $\lambda_t$ starts in an initial value $\lambda_{0}$, $\delta\lambda$ is the amplitude, and $g(t)$ obeys $g(t_0) = 0$ and  $g(t_{f}) = 1$. Thus, $\lambda_t$ varies from $\lambda_0$ to $\lambda_f=\lambda_0+\delta\lambda$.

For small systems, work is a fluctuating quantity \cite{Jarzynski2015} and for a specific protocol $g(t)$ the average work reads
\begin{equation}
\label{eq02}
\la W\ra = \int_{t_{0}}^{t_{f}} \td t \, \dot{\lambda}_t\,\la \pd_\lambda H \ra\, ,
\end{equation} 
where  the angular brackets denote an average over many realizations of the same process and the dot denotes a derivative with respect to time. 

We will now evaluate the general expression for the average work \eqref{eq02} by means of linear response theory. To this end, we expand the Hamiltonian up to linear order in the amplitude $\delta\lambda$,
\begin{equation}
\label{eq03}
H(\lambda_t) = H(\lambda_{0}) + \delta \lambda \, g(t)\, \pd_\lambda H+ \mathcal{O}(\delta\lambda^{2})\,.
\end{equation}	 
By substituting Eq.~\eqref{eq03} into Eq.~\eqref{eq02} and identifying $\pd_\lambda H$ as the generalized force \cite{kub57,kubo2,Bonanca2014a,Acconcia2014}, it can be shown  \cite{Acconcia2014} that the average work \eqref{eq02} becomes
\begin{equation}
\label{eq04}
\begin{split}
\la W\ra  &= \delta\lambda\, \left\langle\pd_\lambda H \right\rangle_{\rho_{0}} - \dfrac{(\delta \lambda)^{2}}{2}  \Psi(0)\\
&- (\delta \lambda)^{2}  \int_{t_{0}}^{t_{f}} \td t \,\pd_t g \int_{0}^{t-t_{0}} \td s \, \Psi(s)\,  \pd_s g(t-s),
\end{split}
\end{equation}
where $\Psi(t)$ is the relaxation function \cite{kub57, kubo2}. 

Until Eq.~\eqref{eq04}, the present treatment is identical to the classical case \cite{Acconcia2014}. However, in the quantum case, the relaxation function $\Psi(t)$ is determined by the quantum response function $\phi(t)$, $\phi(t)=-\dot{\Psi}(t)$, with \cite{kub57,kubo2}
\begin{equation}
\label{eq05}
\phi(t)= \dfrac{1}{i\hbar} \,\tr{\rho_0\,\com{A_0}{A_t}},
\end{equation}  
where $A=\pd_\lambda H$ is the generalized force. To avoid clutter in the formulas, we introduce in Eq.~\eqref{eq05} the notation $A(t)\equiv A_t$.

In complete analogy to the classical case \cite{Acconcia2014}, the first two terms of Eq.~\eqref{eq04} are independent of the specific protocol $g(t)$ and we identify the quasistatic, equilibrium work as
\begin{equation}
\label{eq06}
\la W_\mrm{qs}\ra=\delta\lambda\, \left\langle\pd_\lambda H \right\rangle_{\rho_{0}} - \dfrac{(\delta \lambda)^{2}}{2}  \Psi(0)\,.
\end{equation}
In the remainder of this analysis, we will analyze the excess work,
\begin{equation}
\label{eq07}
\la W_\mrm{ex}\ra=- (\delta \lambda)^{2}  \int_{t_{0}}^{t_{f}} \td t \, \pd_t g\int_{0}^{t-t_{0}} \td s \, \Psi(s)  \,\pd_s g(t-s)
\end{equation}
for two analytically solvable examples. We will show that whenever this thermodynamic quantity vanishes in finite time, the quantum adiabatic invariant is conserved and therefore the system can be driven through a shortcut to adiabaticity.

Generally, it is easy to see that if the adiabatic theorem is fulfilled, no transitions between eigenstates occur, and therefore the excess work $\la W_\mrm{ex}\ra$ has to vanish. However, the reverse is not necessarily true. Even if the excess work vanishes, one could imagine a process during which some transitions between eigenstates do occur, however in such a way that their energetic contribution ``cancels out.'' In the following, we will analyze this issue with the help of two fully analytically solvable examples--the parametric harmonic oscillator and a spin-$1/2$ particle in a magnetic field. We will find that at least within the range of validity of linear response theory, such ``canceling'' transitions do not occur since for a ``shortcut'' not only the excess work vanishes, but also the  adiabatic invariant is (approximately) conserved. For classical systems, a similar analysis was developed in Ref.~\cite{Acconcia2014}.

\section{\label{sec:harm}Parametric harmonic oscillator}

\begin{figure}
\includegraphics[width=.48\textwidth]{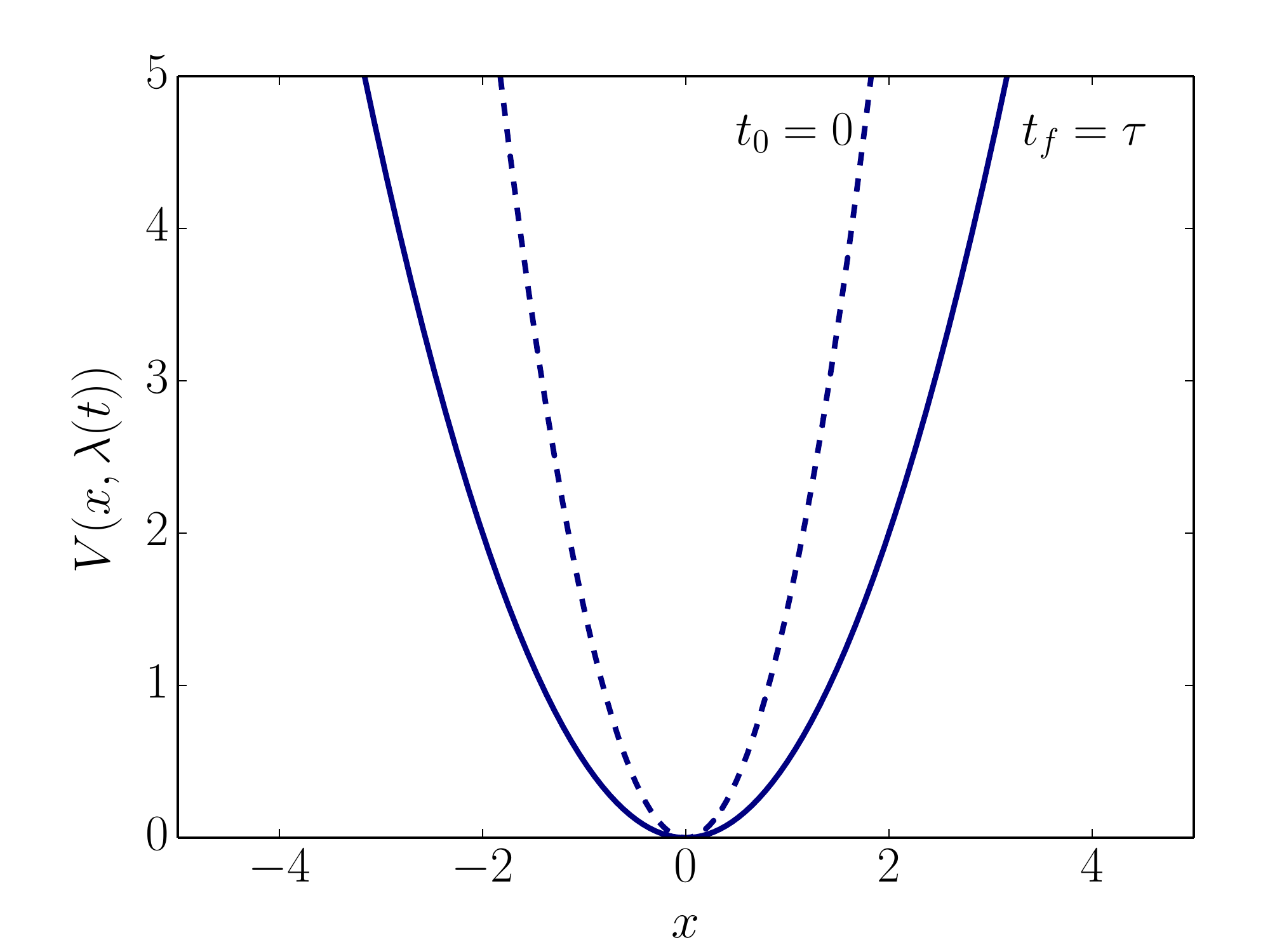} 
\caption{(Color online) Parametric harmonic oscillator \eqref{eq08} with $\lambda_{0}$ (dashed line) at time $t=t_{0}$  and  $\lambda_{f}$ (solid line) at $t=\tau$.}
\label{general_protocol}
\end{figure}

We consider the time-dependent Hamiltonian	
\begin{equation}
\label{eq08}
H(\lambda_t)=\frac{p^2}{2}+\frac{1}{2}\,\lambda_t\,x^2
\end{equation} 
where $x$ and $p$ are the coordinate and momentum operators, respectively. This system can be solved analytically \cite{Deffner2008,Deffner2010} for specific protocols $\lambda_{t}$ that drive the system from an initial to final value of $\lambda$, as illustrated in Fig.~\ref{general_protocol}.
To simplify notation, we further set $t_0=0$ and $t_f=\tau$.

\subsection{\label{sec:lin}Linear response approach}

The response function \eqref{eq05} is obtained by solving Heisenberg's equations of motion for a fixed, initial value of $\lambda$. Hence, we obtain, after a few simple lines,
\begin{equation}
\label{eq10}
\phi(t) = \dfrac{\hbar}{\lambda_0} \, \coth\left(\frac{\beta\hbar\sqrt{\lambda_0}}{2}\right)\,\sin(2\sqrt{\lambda_0}\,t)\,.
\end{equation}
It is interesting to note that the system's response is oscillatory. Consequently, we have the ``relaxation'' function
\begin{equation}
\label{eq11}
\Psi(t) = \dfrac{\hbar}{2\lambda_0\sqrt{\lambda_0}}\,\coth\left(\frac{\beta\hbar\sqrt{\lambda_0}}{2}\right) \, \cos(2\sqrt{\lambda_0}\,t)\,.
\end{equation}
Generally, relaxation functions describe how a system relaxes towards an equilibrium state. However, since the present system has only a single degree of freedom and is thermally isolated, the ``relaxation'' function exhibits nondecreasing oscillations.

For the sake of simplicity, we further assume that the stiffness varies linearly with time,
\begin{equation}
\label{eq12}
\lambda_t=\lambda_0+\delta\lambda\,t/\tau\,,
\end{equation}
 for which we obtain
\begin{equation}
\label{eq13}
\la W_\mrm{ex}\ra = \left(\dfrac{\delta\lambda}{\sqrt{\lambda_0}}\right)^{2} \dfrac{\hbar \sqrt{\lambda_0}}{4} \coth\left(\dfrac{\beta\hbar\sqrt{\lambda_0}}{2}\right) \dfrac{\sin^{2}(\sqrt{\lambda_0}\,\tau)}{\lambda_0\,\tau^{2}}\,.
\end{equation}
Equation~\eqref{eq13} constitutes our first main result. In complete analogy to the classical case \cite{Acconcia2014}, the excess work vanishes for all zeros of the sine function, i.e., for all $\tau=n \pi/\sqrt{\lambda_0}$ with $n$ being an integer. In the classical case, these ``special'' driving times have been attributed to a conservation of the adiabatic invariant during the finite-time process \cite{Acconcia2014}. 

In the next section, we will further analyze this observation and show that the minima of $\la W_\mrm{ex}\ra$ \eqref{eq13}, indeed, identify shortcuts to adiabaticity.

\subsection{\label{sec:exact}Exact solution}

The parametric harmonic oscillator \eqref{eq08} has been extensively studied, since it can be solved analytically \cite{Husimi1953,Deffner2008,Deffner2010,Gong2014} for specific driving protocols and it describes quantum thermodynamic experiments in cold ion traps \cite{Huber2008,Abah2012,Rossnagel2013}. The time-dependent mean energy can be written as \cite{Deffner2010,Deffner2013}
\begin{equation}
\label{eq14}
\la H_\tau\ra=\frac{\hbar\sqrt{\lambda_f}}{2}\,Q^*\,\coth\left(\frac{\beta\hbar\sqrt{\lambda_0}}{2}\right)\,,
\end{equation}
where $Q^*$ is a measure of adiabaticity \cite{Husimi1953,Deffner2008,Deffner2010}. This measure is fully determined by two special solutions, $X_t$ and $Y_t$, of the force-free equation of motion \cite{Husimi1953},
\begin{equation}
\label{eq15}
\ddot{x}_t+\lambda_t\,x_t=0\,.
\end{equation}
We have
\begin{equation}
\label{eq16}
Q^{*} = \dfrac{1}{2\sqrt{\lambda_0\lambda_f}}\left[\lambda_0\left(\lambda_f\,X^{2}_\tau + \dot{X}^{2}_\tau\right) + \left(\lambda_f\,Y^{2}_\tau + \dot{Y}^{2}_\tau\right)\right]\, ,
\end{equation}
with $X_0 = 0$, $\dot{X}_0 = 1$ and $Y_0 = 1$, $\dot{Y}_0 = 0$. \cite{Husimi1953}. Note that these initial conditions for $X_t$ and $Y_t$ are chosen for the sole sake of simplifying the mathematical treatment \cite{Husimi1953}. For the quantum harmonic oscillator, the time-dependent action $S = E(t)/\omega(t)$ is conserved if \cite{Deffner2010}	
\begin{equation}
\label{eq17}
\dfrac{\dot{X}^{2}_t +\lambda_t\,X^{2}_t}{\sqrt{\lambda_t}} = \dfrac{1}{\sqrt{\lambda_{0}}} \quad\mrm{and}\quad \dfrac{\dot{Y}^{2}_t + \lambda_t\,Y^{2}_t}{\sqrt{\lambda_t}} = \sqrt{\lambda_{0}}\,.
\end{equation}
Thus, it is easy to see that $Q^{*} \geq 1$, where the equality holds for quasistatic processes. Accordingly, the exact expression for the excess work reads
\begin{equation}
\label{eq18}
\la W_\mrm{ex}^\mrm{exact}\ra=\frac{\hbar\sqrt{\lambda_0+\delta\lambda}}{2}\coth\left(\dfrac{\beta\hbar\sqrt{\lambda_0}}{2}\right) \,\left(Q^*-1\right)\,.
\end{equation}
Note that $Q^*$ depends only implicitly on the protocol $\lambda_t$ through the solutions of Eq.~\eqref{eq15}. Therefore, it is \textit{ad hoc} not clear whether the exact excess work \eqref{eq18} exhibits the same zeros  as the expression from linear response theory \eqref{eq13} for the linear protocol \eqref{eq12}.

\begin{figure}
\includegraphics[width=.48\textwidth]{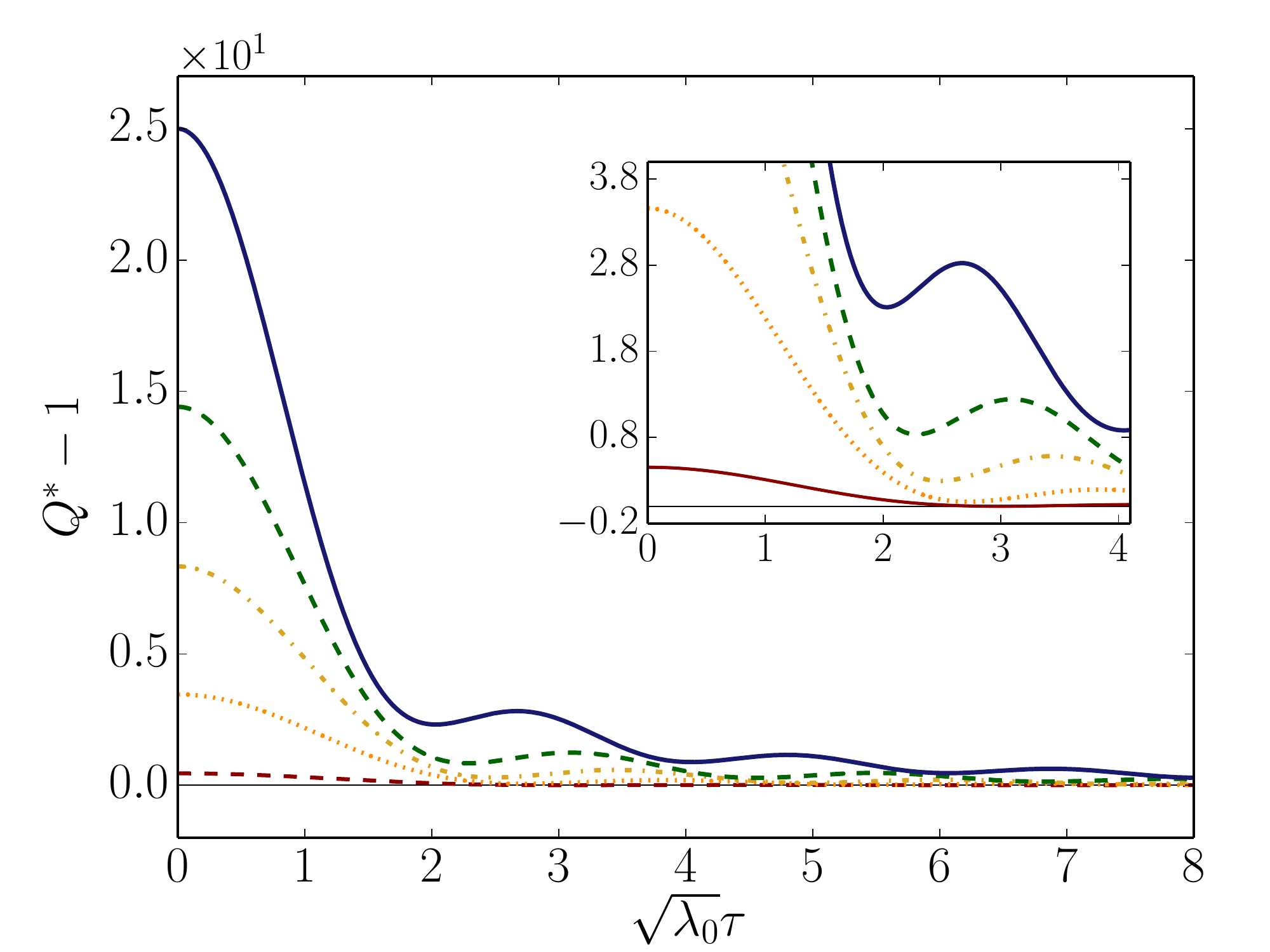} 
\caption{(Color online)  Measure of adiabaticity $Q^*$ \eqref{eq18} as a function of the switching time for the linear protocol, $g(t)=t/\tau$, and $\lambda_f = 2.0$ (blue solid line), $\lambda_f = 1.7$ (green dashed line), $\lambda_f = 1.5$ (yellow dot-dashed line), $\lambda_f = 1.3$ (orange dash-dotted line), and $\lambda_f = 1.1$ (red dotted line), and $\lambda_0 = 1.0$.}
\label{test_omeg1}
\end{figure}

To gain insight and to build intuition, we plot the measure of adiabaticity $Q^*$ \eqref{eq16} for the linear protocol in Fig.~\ref{test_omeg1} for various strengths of the perturbation $\delta\lambda$. We observe that generally $Q^*-1$ exhibits oscillations, but no zeros as a function of $\tau$. 

\begin{figure}
\includegraphics[width=.48\textwidth]{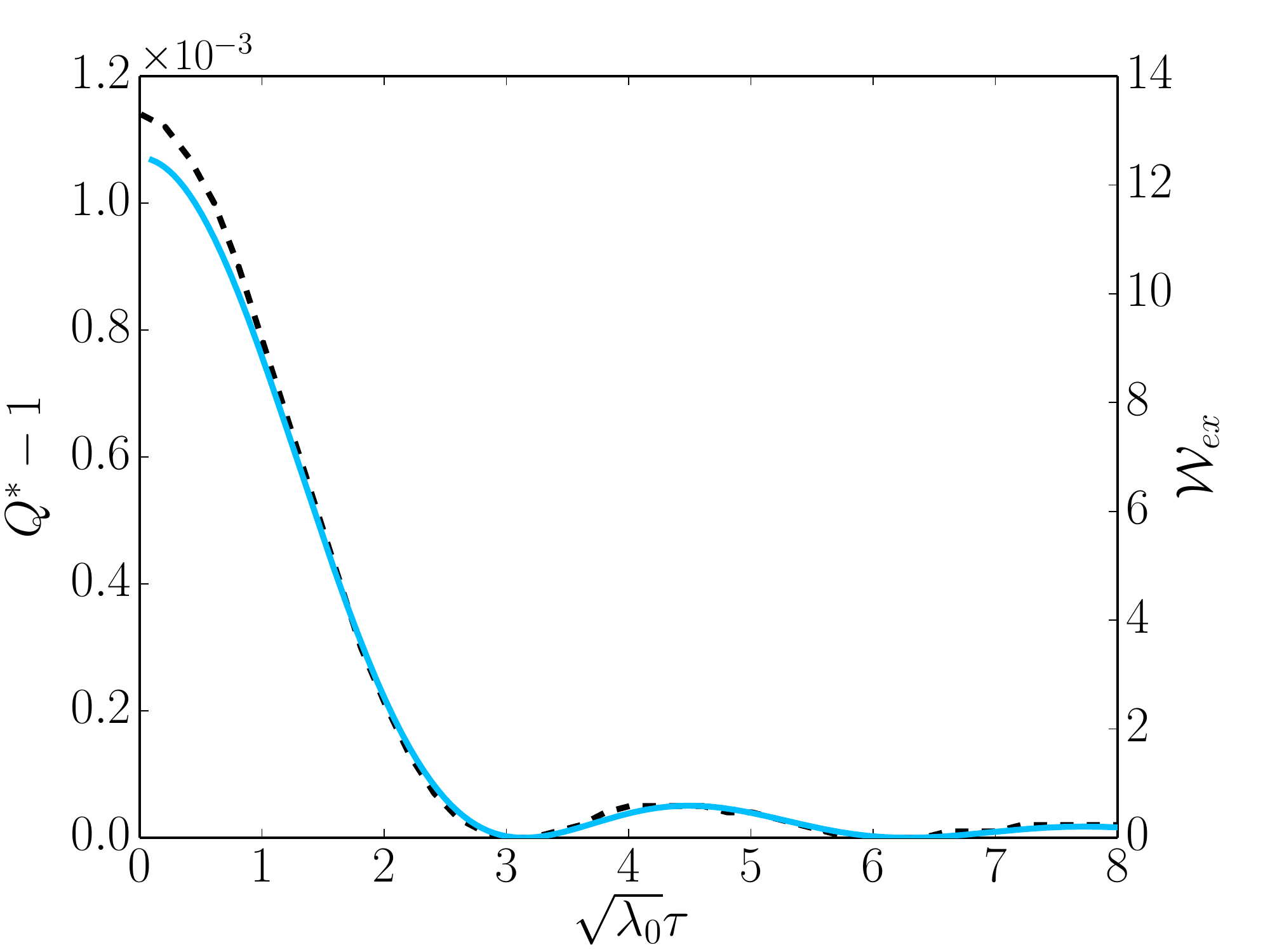} 
\caption{(Color online) Excess work from linear response theory \eqref{eq13} (blue solid line) together with $Q^{*}$-1 (black dashed line) as a function of $\tau$ for the linear protocol \eqref{eq12} and $\delta\lambda = 0.1$. The symbol $\mathcal{W}_{ex}$ denotes $\la W_{ex}\ra$ measured in units of $\left(\delta\lambda/\sqrt{\lambda_0}\right)^{2} (\hbar \sqrt{\lambda_0})\coth\left(\beta\hbar\sqrt{\lambda_0}/2\right)/4$.}
\label{W_exc_linear}
\end{figure}

For weak driving, however, i.e., $\delta\lambda/\lambda_0\ll 1$, where we expect linear response theory to hold, the minima of $Q^*-1$ get infinitely close to zero. In Fig.~\ref{W_exc_linear}, we compare the excess work from linear response theory \eqref{eq13} with the behavior of $Q^*-1$ for weak driving. We observe very good agreement between the result from linear response theory \eqref{eq13} and $Q^*-1$. 

It has also been shown that for $Q^*=1$, the quantum adiabatic theorem is fulfilled, i.e., for such processes there are no transitions between different energy eigenstates \cite{Husimi1953}. Thus, we conclude that the zeros of the excess work, indeed, identify finite driving times for which transitionless quantum driving is facilitated--shortcuts to  adiabaticity from linear response theory.

\subsection{Range of validity of linear response theory}

Linear response theory can be understood as a phenomenological theory of weak perturbations \cite{kub57}. Thus, the numerical and  qualitative agreement between exact \eqref{eq18} and approximate \eqref{eq13} results cannot be considered satisfactory. To deepen the insight into the approximations, we will now derive Eq.~\eqref{eq13} from the exact expression \eqref{eq18} without having to rely on phenomenology.

To this end, we expand the exact expression \eqref{eq18} in powers of $\delta\lambda$ up to second order. Note that $Q^*$ depends implicitly on the protocol $\lambda_t$ and we write $Q^*(\delta\lambda)$. We have
\begin{equation}
\label{eq19}
\begin{split}
\la W_\mrm{exc}^\mrm{exact}\ra&\simeq \frac{\hbar}{2}\coth\left(\dfrac{\beta\hbar\sqrt{\lambda_0}}{2}\right)\,\bigg\lbrace \delta\lambda\,\sqrt{\lambda_0}\,\pd_{\lambda} Q^*(0)\\
&+\frac{\delta\lambda^2}{2\sqrt{\lambda_0}}\left[\pd_{\lambda} Q^*(0)+\lambda_0\, \pd^2_{\lambda} Q^*(0)\right]+\mc{O}(\delta\lambda^3)\bigg\rbrace\,,
\end{split}
\end{equation}
where we used $Q^*(0)=1$. We now have to show that there exist approximate solutions $\mc{X}_t$ and $\mc{Y}_t$ of the equation of motion \eqref{eq15} such that Eq.~\eqref{eq19} reduces to the linear response expression \eqref{eq13} with $\mc{X}_t$ and $\mc{Y}_t$ replacing $X_t$ and $Y_t$ in Eq.~\eqref{eq16}.

Comparing Eqs.~\eqref{eq13} and \eqref{eq19}, we conclude that $\mc{X}_t$ and $\mc{Y}_t$ have to fulfill
\begin{equation}
\label{eq20}
\pd_\lambda Q^*(0)=0\quad\mrm{and}\quad\pd_\lambda^2Q^*(0)=\frac{\sin^2\left(\lambda_0\tau\right)}{\lambda_0^3\tau^2}\,.
\end{equation}
Additionally, we know that $\mc{X}_t$ and $\mc{Y}_t$ have to obey $\dot{\mc{X}}_t\mc{Y}_t - \mc{X}_t\dot{\mc{Y}}_t= 1$ \cite{Husimi1953}. The latter condition is just an expression of the commutation relation between position and momentum \cite{Husimi1953}. For $\delta\lambda=0$, the solution of Eq.~\eqref{eq15} is given by the sine and cosine function \cite{Husimi1953}. Hence, we make the ansatz
\begin{equation}
\label{eq21}
\begin{split}
\mc{X}_t&=\frac{1}{\sqrt{\lambda_0}}\si{\sqrt{\lambda_0} t} +\delta \lambda \,\mc{F}_t+\mc{O}(\delta\lambda^2)\\
\mc{Y}_t&=\co{\sqrt{\lambda_0} t} +\delta \lambda \,\mc{G}_t+\mc{O}(\delta\lambda^2)\,,
\end{split}
\end{equation}
where $\mc{F}_t$ and $\mc{G}_t$ are two time-dependent functions determined by the conditions \eqref{eq20}.

It is then a tedious but straightforward exercise to show that
\begin{equation}
\label{eq22}
\mc{F}_t=\frac{t^2+4 a \lambda_0\tau}{4 \lambda_0\tau}\co{\sqrt{\lambda_0 t}}-\frac{t- 4 b \lambda_0\tau}{4 \lambda_0\sqrt{\lambda_0} \tau}\si{\sqrt{\lambda_0 t}}
\end{equation}
and
\begin{equation}
\label{eq23}
\begin{split}
\mc{G}_t=&-\frac{t^2+4 a \lambda_0\tau}{4 \lambda_0\tau}\co{\sqrt{\lambda_0 t}}\\
&+\frac{t^2\lambda_0-4 c\lambda_0\tau-1}{4 \lambda_0\sqrt{\lambda_0} \tau}\si{\sqrt{\lambda_0 t}}\,.
\end{split}
\end{equation}
The three constants $a$, $b$, and $c$ are determined by the boundary conditions $\mc{F}_0 = a$, $\dot{\mc{F}}_0 = b$ and $\mc{G}_0 = -b$, $\dot{\mc{G}}_0=c$ \footnote{Note that the initial conditions for $\mc{F}$ and $\mc{G}$ are an ansatz chosen to fulfill Eq.~\eqref{eq20}.}. The expressions of $\mc{F}$ and $\mc{G}$ are rather lengthy and can be found in Appendix~\ref{app:lin}.

\begin{figure}
\centering
\includegraphics[width=.45\textwidth]{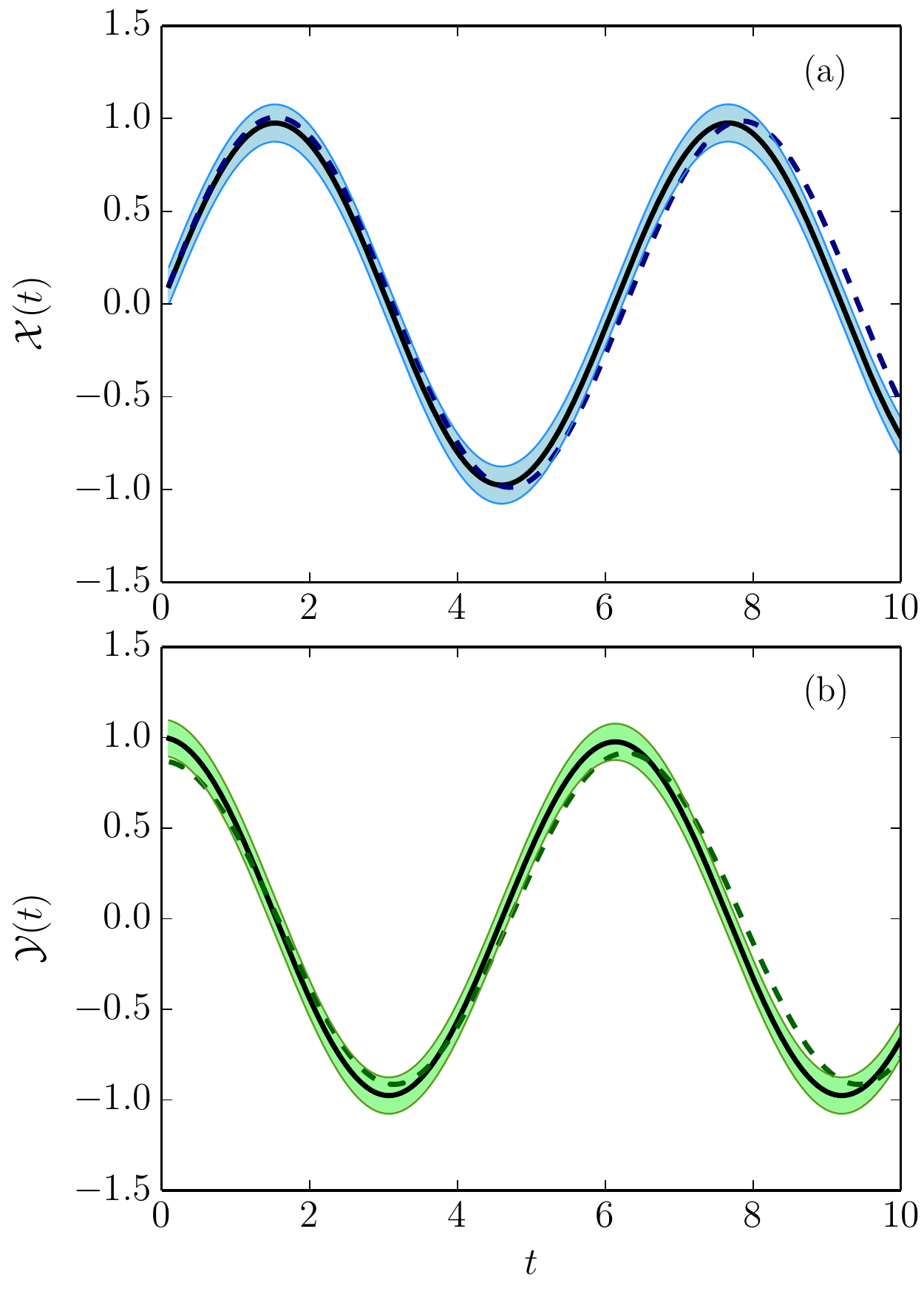} 
\caption{(Color online) (a) Exact solution $X_t$ (solid line) and approximate solution $\mc{X}_t$ \eqref{eq21} for $\delta\lambda = 0.1$ (dashed line). (b) Exact solution $Y_t$ (solid line) and approximate solution $\mc{Y}_t$ \eqref{eq21} for $\delta\lambda$ = 0.1 (dashed line). Shaded area signifies a $\delta\lambda$ environment around the exact results.}
\label{solutions_approx}
\end{figure}

The solutions \eqref{eq21} together with Eqs.~\eqref{eq22} and \eqref{eq23} are the approximate solutions of Eq.~\eqref{eq15}, for which the exact expression for the excess work \eqref{eq18} reduces to the result from linear response theory \eqref{eq13}. In Fig.~\ref{solutions_approx}, we plot the approximate solutions \eqref{eq21} together with the exact solutions of \eqref{eq15}. We observe that $\mc{X}_t$ and $\mc{Y}_t$ are within a $\delta\lambda$ environment around the exact results, as one would intuitively expect by construction.

In conclusion, we have shown that results from linear response theory can also be obtained from expanding the exact solutions for weak driving. Thus, the linear response expressions are not only considered to be qualitatively and phenomenologically true, but also quantitatively exact.

\subsection{\label{sec:deg}Optimal protocols -- shortcuts to adiabaticity}

In an analogous classical treatment, it has been shown that the linear parametrization \eqref{eq12} is not the only protocol with zero excess work. Rather, there is a degenerate family of optimal protocols \cite{seifert,Acconcia2014} for which nonequilibrium excitations are suppressed. This family is given by
\begin{equation}
\label{eq24}
g(t)=t/\tau+ \alpha \ \sin\left( \kappa \pi\,t/\tau\right)\,,
\end{equation}
where $\kappa$ is an integer and $\alpha$ is any arbitrary real number. 

The quantum excess work \eqref{eq13} merely differs in the prefactor from the classical expression
\begin{equation}
\label{eq25}
\la W_\mrm{ex}\ra\bigg|_{\hbar\beta\sqrt{\lambda_0}\ll1} = \left(\dfrac{\delta\lambda}{\sqrt{\lambda_0}}\right)^{2} \dfrac{1}{2\beta} \,\dfrac{\sin^{2}(\sqrt{\lambda_0}\,\tau)}{\lambda_0\,\tau^{2}}\,,
\end{equation}
which is obtained in the limit $\hbar\beta\sqrt{\lambda_0}\ll1$. Thus, the degenerate class \eqref{eq24} constitutes a family of shortcuts to adiabaticity for the quantum harmonic oscillator under weak driving. 
\begin{figure}
\includegraphics[width = 0.46\textwidth]{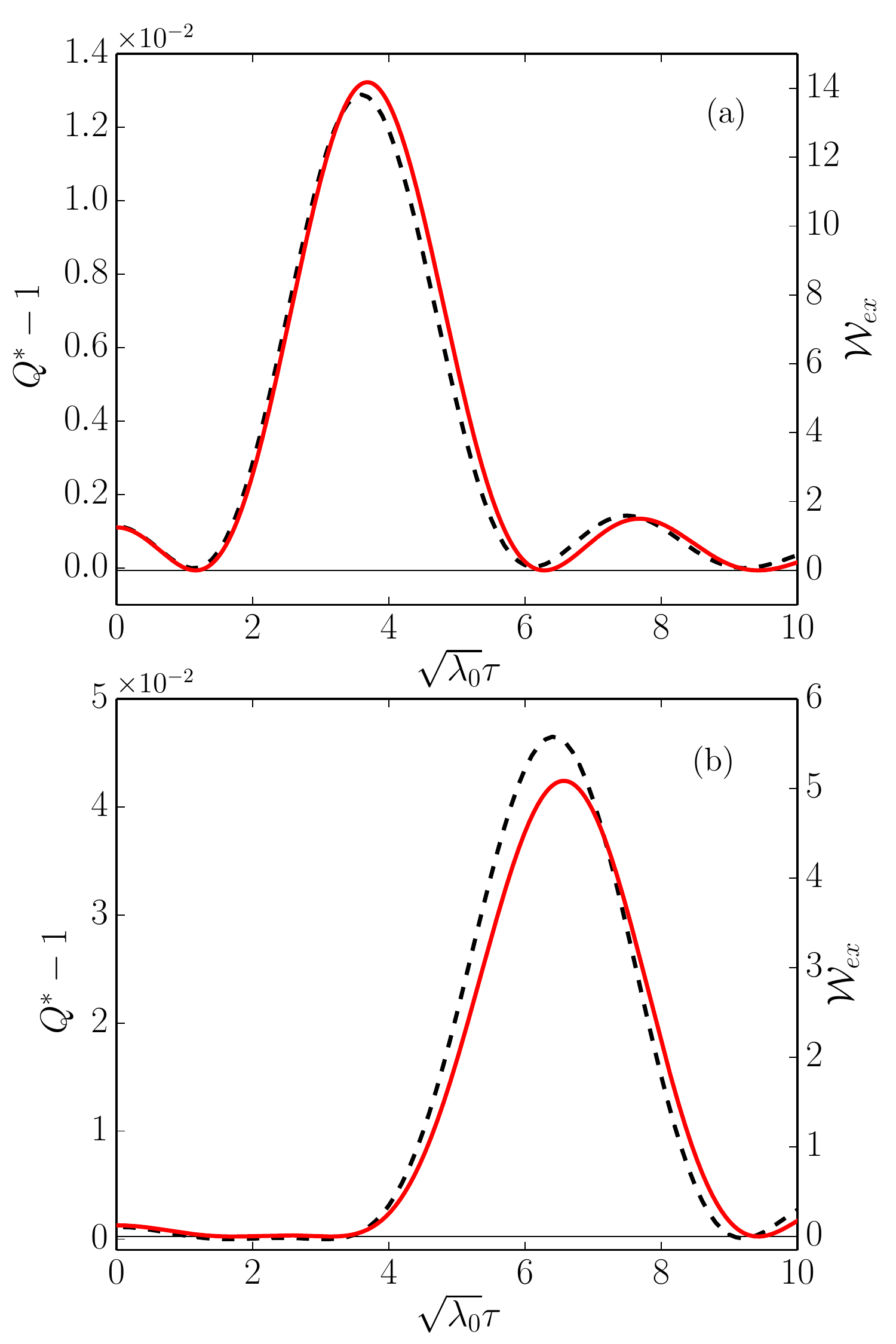} 
\caption{(Color online) Excess work \eqref{eq13} (black dashed line) and normalized adiabatic parameter $Q^{*}-1$ (red solid line) as a function of the switching time for the optimal protocols \eqref{eq24} with (a) $\alpha=1$, $\kappa=2$ and (b) $\alpha=1$, $\kappa=4$. The symbol $\mathcal{W}_{ex}$ denotes $\la W_{ex}\ra$ measured in units of $\left(\delta\lambda/\sqrt{\lambda_0}\right)^{2} (\hbar \sqrt{\lambda_0})\coth\left(\beta\hbar\sqrt{\lambda_0}/2\right)/4$.}
\label{Wexc_sin_protocol}
\end{figure} 
Figure~\ref{Wexc_sin_protocol} illustrates $\la W_\mrm{ex}\ra$ \eqref{eq13} together with $Q^*-1$ for two members of the family \eqref{eq24}. It has been shown \cite{Acconcia2014} that the shortcut to adiabaticity is obtained for $\sqrt{\lambda_{0}}\tau = n\pi$, with $n$ integer, and
\begin{equation}
\label{eq26}
\sqrt{\lambda_0}\tau = \frac{(\kappa\pi/2)}{(1 + \kappa\pi \alpha)^{1/2}}\,.
\end{equation}
Finally, it is worth emphasizing that such shortcuts to adiabaticity can be obtained for arbitrarily short switching times by choosing $\alpha$ appropriately \cite{Acconcia2014}.

\section{\label{sec:spin} Spin 1/2 in a time-dependent magnetic field}

Our second example is a spin 1/2 in a time-dependent magnetic field subjected to the constraint $|\mathbf{B}(t)|=B_{0}=\mathrm{constant}$. Its Hamiltonian reads
\begin{equation}
H(t) = -\frac{\hbar\gamma}{2}\,\boldsymbol\sigma\cdot\mathbf{B}(t)\,,
\label{spinhamil}
\end{equation}
where $\boldsymbol\sigma$ denotes the Pauli matrices. Due to the above-mentioned constraint on $\mathbf{B}(t)$, it is more convenient to choose the following parametrization
\begin{equation}
\mathbf{B}(t) = B_{0}\,\begin{pmatrix}
\sin{[\varphi(t)]} \cos{[\theta(t)]}\\ \sin{[\varphi(t)]} \sin{[\theta(t)]}\\ \cos{[\varphi(t)]}
\end{pmatrix}\,.
\label{magfield}
\end{equation} 
Hence, the time dependence of the set of allowed processes parameterized by the angles $\varphi(t)$ and $\theta(t)$ is, in analogy to Eq.~(\ref{eq01}), expressed as
\begin{subequations}
\begin{align}
\varphi(t) &= \varphi_{0} + \delta\varphi\,g_{\varphi}(t)\,, \\
\theta(t) &= \theta_{0} + \delta\theta\,g_{\theta}(t)\,
\end{align}
\end{subequations}
where the boundary conditions $g_{\varphi, \theta}(0) = 0$ and $g_{\varphi, \theta}(\tau) = 1$ must hold.

Linear response theory provides a good description of $\la W_{ex}\ra$ as long as $\delta\varphi$ and $\delta\theta$ are sufficiently small. In this regime, one can easily show that the angle $\theta(t)$ plays no role and the thermodynamic work (\ref{eq07}) depends on the nonequilibrium of $\partial_{\varphi}H$ only. Thus, the response function is given by Eq.~(\ref{eq05}) with $A_t = \partial_{\varphi}H(t)$ and it is straightforward to obtain
\begin{equation}
\phi(t) = \frac{\hbar}{2}\left(\gamma B_{0}\right)^{2} \tanh{\left( \frac{\beta \hbar \gamma B_{0}}{2}\right)} \sin{\left(\gamma B_{0} t\right)}\,,
\label{respfuncspin}
\end{equation}
from which, using again $\phi(t) = -\dot{\Psi}(t)$, we have the relaxation function
\begin{equation}
\Psi(t) = \frac{\hbar \omega_{0}}{2} \tanh{\left( \frac{\beta \hbar \omega_{0}}{2}\right)} \cos{\left( \omega_{0} t\right)}\,,
\label{eqrelaxspin}
\end{equation}
where we defined $\omega_{0}\equiv \gamma B_{0}$. 

The time dependence of the relaxation functions (\ref{eq11}) and (\ref{eqrelaxspin}) has the same functional form. Therefore, the excess work performed by an external agent while driving the spin 1/2 will behave exactly the same as in the parametric harmonic oscillator. For instance, the protocols given by (\ref{eq24}) also constitute a family of optimal protocols for the present system. Nevertheless, the values of $\omega_{0}\tau$ for which the excess work vanishes are a bit different from those in Fig.~\ref{W_exc_linear} due to the absence of the factor 2 in $\cos{(\omega_{0} t)}$ of Eq.~(\ref{eqrelaxspin}). The linear protocol generates zeros for $\omega_{0}\tau = n\,2\pi$.

\subsection{Quantum adiabatic invariant}

Analogously to Sec.~\ref{sec:harm}, we will now verify that the quantum adiabatic invariant is conserved for spin-1/2 particles driven by Eq.~(\ref{eq24}). To this end, we analyze the time evolution of the coefficients $c_{+}(t)$ and $c_{-}(t)$ appearing in the expansion 
\begin{equation}
|\psi(t) \rangle = \sum_{n = +,-} c_{n}(t) \e{-\frac{i}{\hbar}\int_{0}^{t}\td t'\,E_{n}(t')}\,|n;t \rangle \,,
\end{equation}
of an arbitrary state $|\psi(t)\rangle$. We denote by $E_{n}(t)$ and $|n;t\rangle$ the instantaneous eigenvalues and eigenstates of (\ref{spinhamil}). The quantum adiabatic invariant is then conserved in finite time if, after starting with $c_{+}(0)=1$ and $c_{-}(0)=0$ at the beginning of a certain protocol $g_{\varphi}(t)$, we obtain $c_{+}(\tau) = c_{+}(0)$ and $c_{-}(\tau)=c_{-}(0)$.

\begin{figure}
\centering
\includegraphics[width=0.48\textwidth]{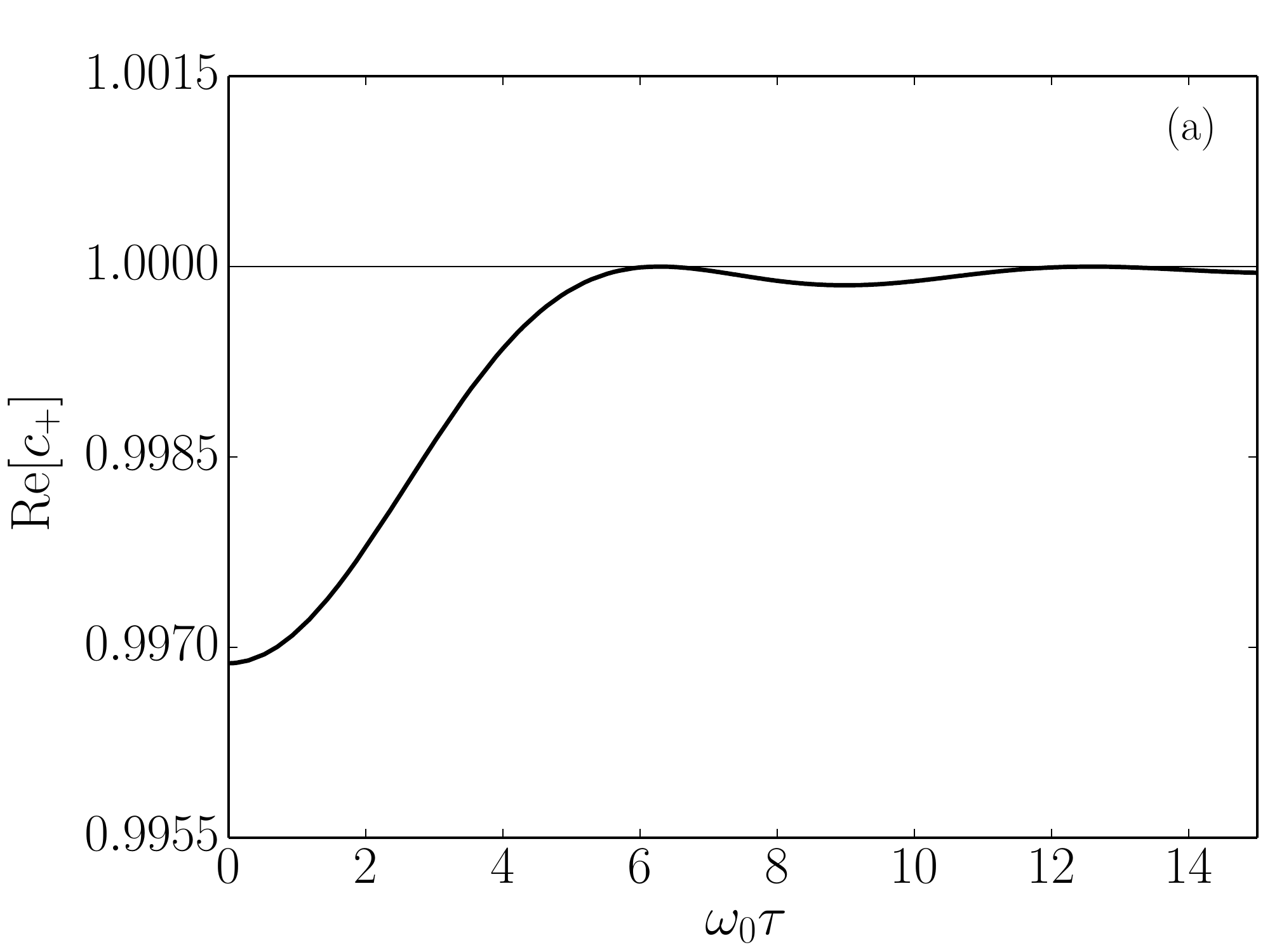}
\includegraphics[width=0.48\textwidth]{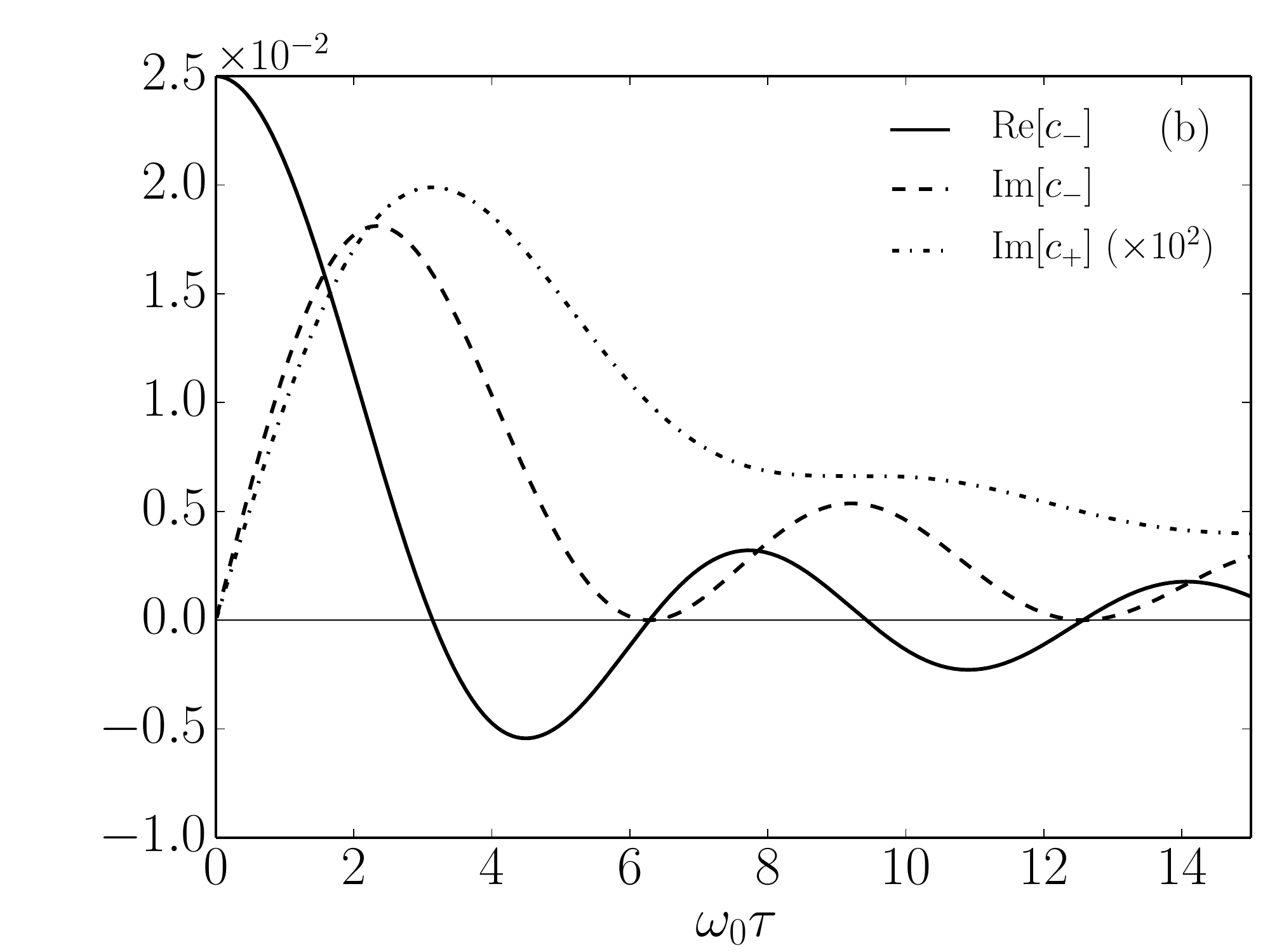}
\caption{Time evolution of the real and imaginary parts of the coefficients $c_{+}(t)$ and $c_{-}(t)$ given by Eq.~(\ref{coeffi}) for the initial condition $c_{+}(0)=1$ and $c_{-}(t)$ for $\cos{(\delta\varphi\, t/2\tau)}\simeq 1$.}
\label{figcoeffi}
\end{figure}

The equations of motion for $c_{+,-}(t)$ are easily derived following standard procedures \cite{Berry1984,Bohm,Pinto2000}. For the parametrization (\ref{magfield}) of $\mathbf{B}(t)$, we obtain
\begin{subequations}
\begin{align}
\dfrac{dc_{+}(t)}{dt} &= - \dfrac{\delta\varphi}{4\tau} \cos\left(\dfrac{\delta\varphi}{2\tau}t\right) \e{-i\omega_{0}t}\, c_{-}(t), \\
\dfrac{dc_{-}(t)}{dt} &=  \dfrac{\delta\varphi}{4\tau} \cos\left(\dfrac{\delta\varphi}{2\tau}t\right) \e{i\omega_{0}t}\, c_{+}(t)\,,
\end{align}
\label{coeffi}
\end{subequations}
considering $\theta_{0}=0$, $g_{\theta}(t)=0$, and $g_{\varphi}(t)=t/\tau$.

Figure \ref{figcoeffi} shows the real and imaginary parts of the solutions of (\ref{coeffi}) as functions of $\omega_{0}\tau$ for $\cos{(\delta\varphi\, t/2\tau)}\simeq 1$, since we are in the regime $\delta\varphi \ll 1$ (see Appendix \ref{app:coeff} for their analytical form). Since the initial conditions are $c_{+}(0)=1$ and $c_{-}(0)=0$, we should have a finite-time conservation of the adiabatic invariant every time we get a recurrence of this values. In Fig.~\ref{figcoeffi}, we see that this holds true for $\omega_{0}\tau=n\, 2\pi$, although due to our approximations the imaginary part of $c_{-}(t)$ does not vanish at these values of $\tau$.

\section{Complex systems}

The two case studies in Secs.~\ref{sec:harm} and \ref{sec:spin} are analytically solvable and pedagogically elucidating. In particular, we obtained exact expressions for the response functions (\ref{eq10}) and (\ref{respfuncspin}). However, this is not feasible  for general and more realistic systems with more degrees of freedom. It has been shown \cite{kub57,kubo2,Sivak,Bonanca2014a} that linear response theory performs well when only phenomenological information is known about the system of interest. In other words, even when the response function is not exact, the predictions of linear response theory provide good approximations.  Finding shortcuts from linear response theory and optimizing $\la W_{ex}\ra$ circumvents the difficult problem of having to solve for the full quantum dynamics. 

\begin{figure}
\centering
\includegraphics[width = 0.48\textwidth]{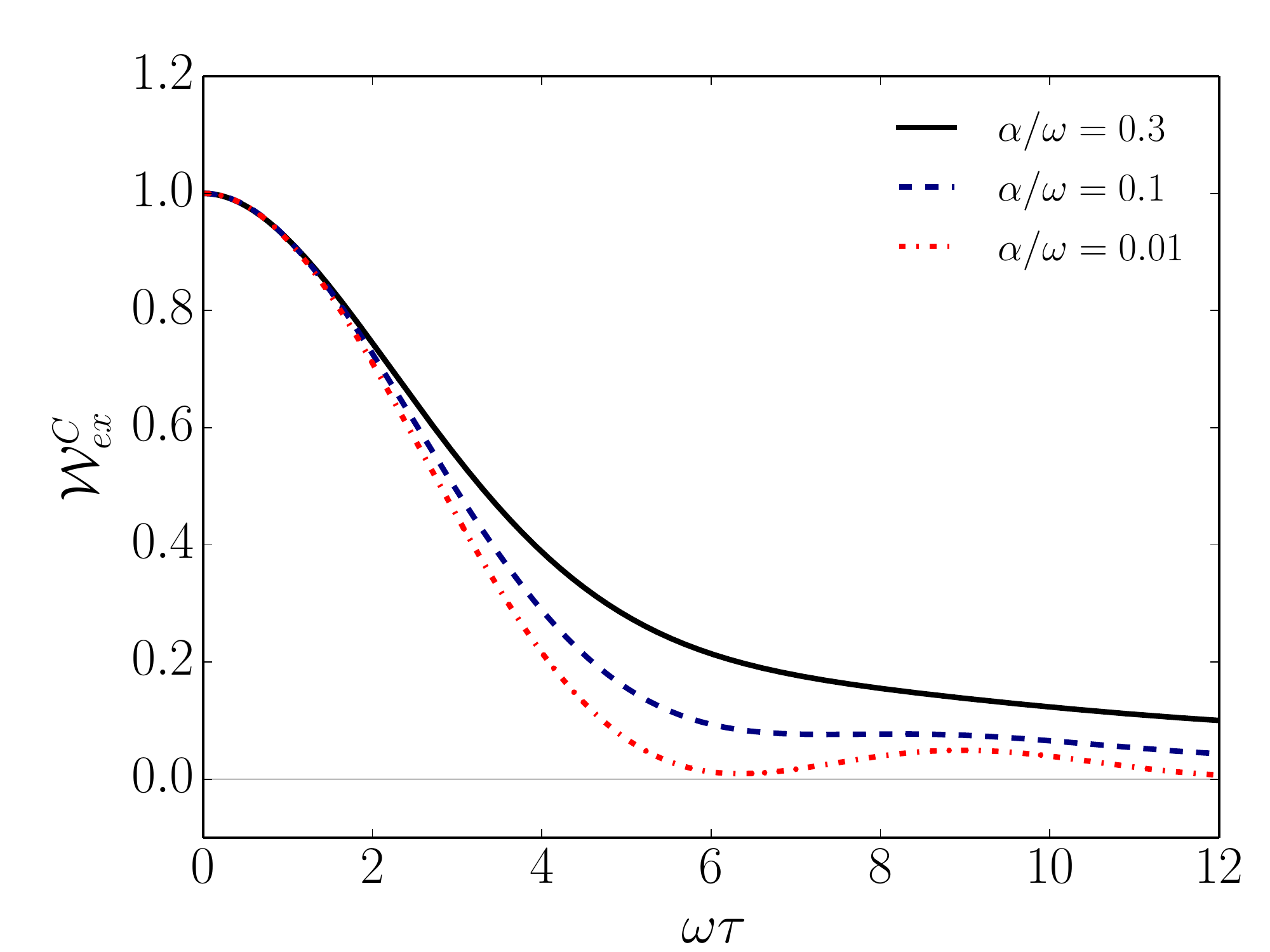} 
\caption{(Color online) Excess work $\mc{W}^C_{ex}$ in units of $(\delta\varphi)^{2}\Psi(0)/2$ for the linear protocol $g_{\varphi}(t)=t/\tau$ and the relaxation function (\ref{phenorelaxfunc}).}
\label{complex}
\end{figure}

It has been shown that $\la W_{ex}\ra$ will have finite-time minima, or nonmonotonic behavior as a function of $\tau$, if the relaxation function is sufficiently oscillatory. This can be illustrated, for instance, using the following phenomenological ansatz \cite{kubo2,Bonanca2014a}:
\begin{equation}
\Psi(t) = \Psi(0)\, \e{-\alpha |t|}\,\left( \cos{(\omega t)} + \frac{\alpha}{\omega} \sin{(\omega t)}\right)\,,
\label{phenorelaxfunc}
\end{equation}
for the relaxation function. Plugging the expression above in Eq.~(\ref{eq07}), we obtain the results shown in Fig.~\ref{complex} for different values of $\alpha/\omega$. As this ratio decreases, the excess work starts to show minima whose value approaches zero. There are several systems for which Eq.~(\ref{phenorelaxfunc}) describes  the relaxation dynamics very well. Among them, we mention a system composed of weakly interacting magnetic moments in the regime where Bloch equations are valid \cite{white}.

\vspace{1em}

\section{Concluding Remarks}

Identifying optimal quantum processes with suppressed or even vanishing nonequilibrium excitations is an important topic, which has recently been attracting intense research efforts. However, all methods currently available necessitate the solution of the full quantum dynamics. In the present work, we have proposed a phenomenological alternative. By generalizing our previous result for the excess work from linear response theory to a quantum system, we have shown that shortcuts to adiabaticity can be identified from a mathematically simple theory. This observation has been proven for two paradigmatic examples of quantum thermodynamics, namely the parametric harmonic oscillator and the spin 1/2 in a time-dependent magnetic field.

\begin{acknowledgments}
T.A. acknowledges support from the Institute of Physics Gleb Wataghin, CNPq (Brazil), Project No. 134296/2013-3, and Capes (Brazil), Project No. 1504869. M.B. acknowledges financial support from FAPESP (Brazil), Project No. 2012/07429-0. S.D. acknowledges financial support from the U.S. Department of Energy through a LANL Director's Funded Fellowship. 
\end{acknowledgments}

\appendix

\begin{widetext}
\section{Approximate solution within linear response\label{app:lin}}

The full expressions for the approximate solutions $\mc{X}_t$ and $\mc{Y}_t$ in Eqs.~\eqref{eq22} and \eqref{eq23} are given in terms of the three constants $a$, $b$, and $c$. These are determined by solving the force-free equation of motion \eqref{eq15} with the boundary conditions $\mc{F}_0 = a$, $\dot{\mc{F}}_0 = b$ and $\mc{G}_0 = -b$, $\dot{\mc{G}}_0=c$. We have
\begin{equation}
\label{eqa1}
a = \dfrac{-1-2\lambda_{0}\tau^{2} + \cos(2\sqrt{\lambda_{0}}\tau) - 2\sqrt{\lambda_{0}}\tau \sin(2\sqrt{\lambda_{0}}\tau)}{8 \lambda_{0}^{2} \tau}\quad \mrm{and}\quad c =  \dfrac{-1+2\lambda_{0}\tau^{2} + \cos(2\sqrt{\lambda_{0}}\tau) - 2\sqrt{\lambda_{0}}\tau \sin(2\sqrt{\lambda_{0}}\tau)}{8 \lambda_{0} \tau} 
\end{equation}
and
\begin{equation}
\label{eqa2}
b = \frac{1}{8 \lambda_{0}\sqrt{\lambda_0}\tau}\left[ 2 + \sqrt{4+2\lambda_{0}\tau^{2} - 4 \cos(2\sqrt{\lambda_{0}}\tau) }  + 2 \sqrt{\lambda_{0}}\tau \cos(2\sqrt{\lambda_{0}}\tau) + \sin(2\sqrt{\lambda_{0}}\tau)\right] \,.
\end{equation}

\section{Derivation of time-dependent coefficients $c_{+}$ and $c_{-}$\label{app:coeff}}

According to Ref.~\cite{Pinto2000}, the coefficients $c_{+}$ and $c_{-}$ satisfy the differential equations for Eq.~(\ref{magfield}),
\begin{subequations}
\begin{align}
\dfrac{dc_{+}(t)}{dt} &= - \dfrac{\delta\varphi}{4\tau} \cos\left(\dfrac{\delta\varphi}{2\tau}t\right)\, \e{-i\omega_{0}t} c_{-}(t), \\
\dfrac{dc_{-}(t)}{dt} &=  \dfrac{\delta\varphi}{4\tau} \cos\left(\dfrac{\delta\varphi}{2\tau}t\right)\, \e{i\omega_{0}t} c_{+}(t).
\end{align}
\end{subequations}
In the regime $\delta\varphi \ll 1$, we make the approximation $\cos[\delta\varphi/2\tau\, t] \simeq 1$. Next, we solve exactly the equations with the initial conditions $c_{+}(0) = 1$ and $c_{-}(0) = 0$. After making $t=\tau$, we obtain the following equations:
\begin{eqnarray}
c_{+}(\omega_{0}\tau) &=& \dfrac{e^{-\frac{i \omega_{0}\tau}{2}}}{\delta\varphi^{2} + 4 (\omega_{0}\tau)^{2}} \left[ (\delta\varphi^{2} + 4 (\omega_{0}\tau)^{2}) \cosh\left(\dfrac{1}{4}\sqrt{-(\delta\varphi^{2} + 4 (\omega_{0}\tau)^{2})}\right)\right. \nonumber \\ 
&-&  \left. 2 i \omega_{0}\tau \sqrt{-(\delta\varphi^{2} + 4 (\omega_{0}\tau)^{2})}  \sinh\left( \dfrac{1}{4}\sqrt{-(\delta\varphi^{2} + 4 (\omega_{0}\tau)^{2})} \right) \right]
\label{c+}
\end{eqnarray}
\begin{equation}
c_{-}(\omega_{0}\tau) =  \dfrac{e^{\frac{i \omega_{0}\tau}{2}} \delta\varphi}{\delta\phi^{2} + 4 (\omega_{0}\tau)^{2}} \sinh\left( \dfrac{1}{4}\sqrt{-(\delta\varphi^{2} + 4 (\omega_{0}\tau)^{2})} \right)\,,
\label{c-}
\end{equation}
where $\omega_{0} = \gamma B$.
\end{widetext}

\bibliography{paper_Quantum2}

\end{document}